\input harvmac
\input epsf
\overfullrule=0pt
\Title{\vbox{
\hbox{USC-97/10}
\hbox{hep-th/9806246}
}}
{\vbox{\centerline{Duality Symmetries for $N=2$ Supersymmetric QCD}
\medskip\centerline{with Vanishing $\beta$-functions}}}{\baselineskip=12pt
\centerline{Joseph A. Minahan}
\bigskip
\centerline{\sl  Department of Physics and Astronomy}
\centerline{\sl University of Southern California}
\centerline{\sl Los Angeles, CA 90089-0484}
\medskip
\bigskip
\centerline{\bf Abstract}

\def\cT{{\cal T}}
\def\cS{{\cal S}}

We construct the duality groups for $N=2$ Supersymmetric QCD with 
gauge group $SU(2n+1)$ and $N_f=4n+2$ hypermultiplets in the fundamental
representation.  The groups are generated by two elements $\cS$ and $\cT$ that
satisfy a relation $\left(\cS\cT\cS^{-1}\cT\right)^{2n+1}=1$.  Thus,
the groups are not subgroups of $SL(2,Z)$.   We
also construct automorphic functions that map the fundamental region of the 
group generated by $\cT$ and $\cS\cT\cS$ to the Riemann sphere.  These
automorphic functions faithfully represent the duality symmetry in
the Seiberg-Witten curve.}

\noindent

\Date{October 1997}
\vfil
\eject

\def\nup#1({Nucl.\ Phys.\ $\us {B#1}$\ (}
\def\plt#1({Phys.\ Lett.\ $\us  {B#1}$\ (}
\def\cmp#1({Comm.\ Math.\ Phys.\ $\us  {#1}$\ (}
\def\prp#1({Phys.\ Rep.\ $\us  {#1}$\ (}
\def\prl#1({Phys.\ Rev.\ Lett.\ $\us  {#1}$\ (}
\def\prv#1({Phys.\ Rev.\ $\us  {#1}$\ (}
\def\mpl#1({Mod.\ Phys.\ Let.\ $\us  {A#1}$\ (}
\def\ijmp#1({Int.\ J.\ Mod.\ Phys.\ $\us{A#1}$\ (}
\def\jag#1({Jour.\ Alg.\ Geom.\ $\us {#1}$\ (}
\def\us#1{\underline{#1}}

\def\NP{{\it Nucl. Phys.\ }}
\def\PL{{\it Phys. Lett.\ }}

\def\PRL{{\it Phys. Rev. Lett.\ }}

\def\cT{{\cal T}}
\def\cS{{\cal S}}
\def\cV{{\cal V}}
\def\tt{\tilde\tau}
\def\al{\alpha}
\def\la{\lambda}
\def\Re{{\rm Re}}
\def\Im{{\rm Im}}
\def\cW{{\cal W}}
\def\CD{{\cal D}}
\def\CC{{C}}
\def\inbar{\,\vrule height1.5ex width.4pt depth0pt}
\def\IC{{\relax\hbox{$\inbar\kern-.3em{\rm C}$}}}

\lref\Harvey{J.~A.~Harvey, {\it Magnetic Monopoles, Duality, and
Supersymmetry}, Trieste HEP Cosmology 1995:66-125, {\bf hep-th/9603086};
J.~ Gauntlett, {\it Duality and Supersymmetric Monopoles}, hep-th/9705025.}
\lref\SWI{N. Seiberg and E. Witten, {\bf hep-th/9407087},
        \NP{\bf 426} (1994) {19}.}
\lref\SWII{N. Seiberg and E. Witten, {\bf hep-th/9408099},
        \NP {\bf B431} (1994) {484}.}
\lref\KLTY{A. Klemm, W. Lerche, S. Theisen and S. Yankielowicz,
        {\bf hep-th/9411048}, \PL {\bf B344} (1995) {169}.}
\lref\AF{P.C. Argyres and A.E. Faraggi, {\bf hep-th/9411057},
        \PRL {\bf 73} (1995) {3931}.}
\lref\HO{A. Hanany and Y. Oz, {\it On the
	Quantum Moduli Space of Vacua of N=2 Supersymmetric $SU(N_c)$
	Gauge Theories, {\bf hep-th/9505075}, {\it Nucl. Phys.} {\bf B452} 
(1995) 283-312.}}
\lref\APS{P.~C.~Argyres, M.~R.~Plesser and A.~D.~Shapere, 
{\it The Coulomb Phase of N=2 Supersymmetric QCD}, {\bf hep-th/9505100},
{\it Phys. Rev. Lett.} {\bf 75} (1995) 1699-1702.} 
\lref\FK{H.~M.~Farkas and I.~Kra, {\it Riemann Surfaces}, Springer-Verlag 
(1980), New York.}
\lref\Clemens{C.~H.~Clemens, {\it A Scrapbook of Complex Curve Theory}, Plenum
Press (1980), New York.}
\lref\MNI{J. Minahan and D. Nemeschansky,  
{\it Hyperelliptic curves for Supersymmetric Yang-Mills}, 
{\it Nucl.Phys.} {\bf B464} (1996) 3,
{\bf hep-th/9507032}.}
\lref\MNII{J. Minahan and D. Nemeschansky, 
{\it N=2 Super Yang-Mills  and Subgroups of $SL(2,Z)$,}
{\it Nucl.Phys.} {\bf B468} (1996) 72,{\bf hep-th/9601059}.}
\lref\AD{P. Argyres and M. Douglas, {\it New Phenomena in $SU(3)$ 
Supersymmetric Gauge Theory}, {\bf hep-th/9505062}, 
{\it Nucl. Phys.} {\bf B448} (1995) 93-126}
\lref\Argyres{P.~Argyres, {\it S-Duality and Global Symmetries in $N=2$
 Supersymmetric Field Theory}, {\bf hep-th/9706095}.}
\lref\ArgyresII{P.~Argyres, {\it Dualities in Supersymmetric Field Theories}, {\bf hep-th/9705076}.}
\lref\WitM{E.~Witten, {\it Solutions Of Four-Dimensional Field Theories Via 
M Theory}, {\bf hep-th/9703166}.}
\lref\AY{O. Aharony and S. Yankielowicz, {\it Exact Electric-Magnetic Duality 
in $N=2$ Supersymmetric QCD Theories}, {\it Nucl.Phys.} {\bf B473} (1996) 
93, {\bf hep-th/9601011}.}
\lref\Koblitz{N. Koblitz, {\it Introduction to Elliptic Curves and 
Modular Forms},
Springer-Verlag, (1984), New York.}
\lref\Sch{B. Schoeneberg, {\it Elliptic Modular Functions; An Introduction},
Springer-Verlag, (1974), New York.}
\lref\FP{D. Finnell and P. Pouliot, {\it Instanton Calculations Versus Exact 
Results In 4 Dimensional Susy Gauge Theories}, {\bf hep-th/9503115}.}  
\lref\DW{R. Donagi and E. Witten, {\it Supersymmetric Yang-Mills Systems And 
Integrable Systems}, {\bf  hep-th/9510101}.}
\lref\MWI{E. Martinec and N. Warner, {\it Integrable systems and 
supersymmetric gauge theory}, {\bf hep-th/9509161}.}
\lref\Mart{E. Martinec, {\it Integrable Structures in Supersymmetric Gauge 
and String Theory}, {\bf hep-th/9510204}.}
\lref\MWII{E. Martinec and N. Warner, {\it Integrability in N=2 Gauge Theory: 
A Proof}, {\bf hep-th/9511052}.}
\lref\GKMMM{A.Gorsky, I.Krichever, A.Marshakov, A.Mironov and A.Morozov,
{\it Integrability and Seiberg-Witten Exact Solution}, {\bf hep-th/9505035},
{\it Phys. Lett.} {\bf B355} (1995) 466-47.}
\lref\NT{T. Nakatsu and K. Takasaki, {\it Whitham-Toda hierarchy and $N = 2$
 supersymmetric Yang-Mills theory}, {\bf hep-th/9509162}.}
\lref\IM{H. Itoyama and A. Morozov, {\it Integrability and Seiberg-Witten 
Theory: Curves and Period}, {\bf hep-th/9511126}.}
\lref\Lehner{J.~Lehner, {\it Discontinous Groups and Automorphic Functions},
{\it American Mathematical Society}, 1964.}
\lref\DKM{N.~Dorey, V.~ Khoze and  M.~Mattis, \plt{388} (1996) 324, 
hep-th/9607066; \nup{492} (1997) 607, hep-th/9611016}
\lref\AB{P.~Argyres and A.~Buchel, {\it The Nonperturbative Gauge Coupling of 
$N=2$ Supersymmetric Theories}, hep-th/9806234.}

\newsec{Introduction}

It is widely believed, and there is a large amount of evidence to support
this, that $N=4$ Yang-Mills has a nontrivial duality symmetry
(see \Harvey\ for reviews and references therein).  
Given the coupling $\tau={\theta\over2\pi}+
{4\pi i\over g^2}$,  it is thought that the transformation 
$\tau\to -~1/\tau$
is an exact symmetry if one also exchanges the root lattice with the
coroot lattice.

It has also been argued that there are duality symmetries for
$N=2$ theories if the $\beta$ function vanishes\refs{\SWII\HO\APS\MNI\AY{--}
\MNII}.  
In this case,
the theory has a dimensionless parameter, which is more or less
the bare coupling up to some quantum corrections.  The most famous
example of this is the $SU(2)$, $N_f=4$ theory, where the duality
properties have been worked out in great detail\SWII.

Less clear are the duality properties for the higher gauge groups.
Recently, it has been argued that the duality group is isomorphic
to the $SL(2,Z)$ subgroup $\Gamma_0(2)$ \refs{\WitM,\Argyres}.  
$\Gamma_0(2)$ is generated
by two elements, $S$ and $T^2$, where $T^2$ is the transformation
that shifts the $\theta$-angle by $2\pi$ and $S$ is the $Z_2$ transformation
that takes weak to strong coupling.  In a very elegant argument,
Argyres demonstrated that this $Z_2$ symmetry is an exact symmetry
for the full theory, by relating it to a parity symmetry of a larger,
asymptotically free gauge theory\Argyres.

From the perspective of $M$ theory\WitM,  there is a function $w$,
\eqn\weq{
w=- {4\la_+\la_-\over(\la_+ - \la_-)^2},
}
where $\la_+$ and $\la_-$ are related to the positions of two five branes
out at infinity.  $w$ is invariant under the interchange of $\la_+$ with
$\la_-$ or $\la_+$ with $-\la_-$.  This second transformation has a fixed
point at $\la_+=-\la_-$, where $w=1$.  This is a $Z_2$ orbifold point.
There is also a weak coupling point at $w=0$ and a strong coupling point
at $w=\infty$.  As a map from the upper half $\tau$ plane to
the sphere, $w$ has nontrivial monodromy at the points $w=0,1$ and $w=\infty$.
Since $w=0$ is a weak coupling point, the generator of the monodromy about
this point is an element $\cT$, which shifts the $\theta$-angle by $2\pi$.
The generator of the monodromy about $w=1$ is an element $\cS$, where
$\cS^2=1$. Therefore, the generator of the monodromy about $w=\infty$
is $\cS\cT$.  At this point an implicit assumption is made that there
does not exist a nonzero integer $n$ such that $(\cS\cT)^n=1$.  Without
any such relation, the duality group is isomorphic to $\Gamma_0(2)$.

But nothing
in $M$ theory explicitly prohibits the presence of such a relation 
and it is the purpose of this paper to show that such relations
do exist, with the relation depending on the gauge group in question.  
A careful analysis
shows that  for $SU(n)$ with odd $n$ and $N_f=2n$ there exisits the
relation
\eqn\STrel{
(\cS\cT\cS^{-1}\cT)^n=1.
}
Therefore, the duality group is not isomorphic to $\Gamma_0(2)$.

In section 2 we make some preliminary remarks about duality symmetries.
In section 3 we discuss the special case of $SU(3)$ with $N_f=6$.  In
this case the duality group, $\Gamma$, has a subgroup $\Gamma'$ which
 is isomorphic to $\Gamma_1(3)$.
In section 4 we discuss the duality group for general $SU(n)$ with
$N_f=2n$.  For these cases, the duality groups do not have nonabelian
subgroups that are isomorphic to subgroups of $SL(2,Z)$.  In section 5
we discuss the automorphic forms that arise for these duality groups.

\newsec{Some Duality Preliminaries}

Before continuing, we should define what we mean by duality.  In an
$N=4$ theory one can describe the duality as follows \ArgyresII:  
Classically the
physics is determined by a coupling parameter $\tau$ that lives in the
upper half plane.  When the theory is quantized, there are no quantum 
corrections to the $\tau$ parameter, but one finds that  different
regions of the upper half plane are identified under $SL(2,Z)$
transformations.  Under such transformations, vector bosons, 
monopoles and dyons are mapped into each other.  Thus, we can describe
the duality by either showing how the regions of the upper half plane are 
identified, or by 
detailing the transformations of the charged states under the duality group.

In an $N=2$ theory with vanishing $\beta$-function the notion of duality is 
a little more subtle.  One 
significant difference is that the classical coupling {\it does} receive
quantum corrections, even when the masses of the hypermultiplets are zero.
So it is not even clear if a classical $\tau$ parameter is  well defined 
beyond weak coupling.  One can define a quantum $\tau$ parameter, but
then the question arises of which one should be used.  In the $SU(2)$ case,
the natural $\tau$ parameter is the effective coupling for the massless theory 
\refs{\MNII,\DKM}.  Under $SL(2,Z)$ transformations of this $\tau$ parameter
the theory is mapped into itself.  For higher $SU(n)$, the effective coupling
is a matrix, so picking a $\tau$ parameter is much less straightforward.

Let us assume that there is a parameter $\tau$ that we
can use for a given $SU(n)$ theory with $N_f=2n$ massless hypermultiplets.  The
Seiberg-Witten hyperelliptic curve for this theory is \APS
\eqn\precurve{
y^2=(P(x))^2-(1-h^2(\tau))x^{2n}\qquad\qquad 
P(x)=x^n-\sum_{j=2}^n u_j x^{n-j}
}
where the $u_j$ are the $SU(n)$ invariants and $h(\tau)$ is some
function that behaves as $1-h(\tau)\sim e^{2\pi i\tau}$ for large imaginary
$\tau$.  At weak coupling, the classical coupling is 
\eqn\class{
\tau_{cl}={1\over2\pi i}\log(h^2(\tau)-1).
}  
Since $h(\tau)$ can presumably take on all possible
values, this definition for $\tau_{cl}$ will break down at strong
coupling because the imaginary part of the right hand side can be negative.

The Seiberg-Witten differential for \precurve\ is \APS
\eqn\SWdiff{
d\lambda~=~{x(ydP-Pdy)\over P^2-y^2}={n(1-h^2)x^{2n}Pdx-(1-h^2)x^{2n+1}dP\over
y(1-h^2)x^{2n}}=-\sum_{j=2}^n j u_j {x^{n-j}dx\over y}
}
Hence, $d\lambda$ is a linear combination of the holomorphic differentials
$\omega_i$ of the curve in \precurve.  The Seiberg-Witten
coordinates $\phi_i$ are given by integrals of $d\lambda$ around the
$a$ cycles and the dual coordinates are given by integrals of $d\lambda$
around the $b$ cycles.  We can define a basis of $a$ and $b$ cycles
by choosing
\eqn\abcycle{
\oint_{a_i}\omega_j=\delta_{ij}\qquad\qquad\oint_{b_i}\omega_j=\Omega_{ij},
}
where $\Omega_{ij}$ is the period matrix.  Therefore, the dual coordinates
satisfy
\eqn\dualvar{
\phi_{Di}=\Omega_{ij}\phi_j.
}
The $Sp(2n-2,Z)$ transformations that act on the
period matrix naturally act on the BPS states.  Hence, we expect the massless
theory to have a nontrivial duality symmetry.  Although $\Omega_{ij}$ is 
positive imaginary, there are still curves of marginal
stability.   But it is quite possible that the positivity of
${\rm Im}\Omega_{ij}$
will play an important role in the stability of BPS states.

The $Sp(2n-2,Z)$ transformations are generated either by monodromies in
the $u_j$ complex planes, or in the $h$ complex plane.  The monodromies
of the first type generate Weyl reflections with quantum corrections.  
However, a rotation of $u_n$ outside the singularities will have no
quantum corrections.  The $Sp(2n-2,Z)$ matrices
that correspond to Weyl reflections have the form 
$\left(\matrix{A&0\cr0&(A^{T})^{-1}}\right)$ where $A$ is a $n-1$ by
$n-1$ matrix that is an element of the group of permutations on $n$ objects
for the irreducible representation of order $n-1$.

The monodromies in the $h$ complex plane  generate the duality group.  
The function
$h(\tau)$ is related to the $M$ theory function $w$ by $w=1-1/h^2$.
Hence, there is nontrivial monodromy about $h^2=0,1,\infty$.  
The normalization of $\tau$ is such that $h(\tau)$ is invariant under the
shift $\tau\to \tau+1$.  This is the transformation that shifts the
$\theta$-angle by $2\pi$, and hence is a symmetry of the theory.  We
call the generator of this transformation $\cT$. This is the monodromy
about $h=\pm1$. 

Since the curve \precurve\ is invariant under 
$h(\tau)\to -h(\tau)$, there is a $Z_2$ monodromy about $h=\infty$.  
If we were to assume that 
\eqn\heqII{
h(\tau)={{\vartheta_3}^4(2\tau)\over
{\vartheta_4}^4(2\tau)-{\vartheta_2}^4(2\tau)},
}
 which is the 
function normally used for the $SU(2)$ gauge 
group\foot{The factor of 2 appears in the
$\vartheta$-function arguments because of our normalization
of $\tau$}, then the transformation
$\tau\to -1/(4\tau)$ sends $h(\tau)$ to $-h(\tau)$.  Hence we
will call this transformation $\cS$.

The transformations $\cS$ and $\cT$ generate the duality group.  
Obviously, there is the relation $\cS^2=1$.  
If there is no other relation then the
group is isomorphic to $\Gamma_0(2)$.  If $h(\tau)$ is the function
in \heqII, then as far as its action on $\tau$ is concerned, there
is no other relation.

However, the duality group is established by 
the action of $\cS$ and
$\cT$ on the BPS states in the theory.  
This means that we should determine
the action of $\cS$ and $\cT$ on $\Omega$ and look
for combinations of $\cS$ and $\cT$  that leave it fixed.  One can do this
without explicity knowing the function
for $h(\tau)$. If there is a
combination that leaves $\Omega$ invariant (up to
a Weyl reflection), then the
duality group is not $\Gamma_0(2)$.  

If the group is not $\Gamma_0(2)$, then
it is somewhat unsatisfying to use the function in \heqII\ to define
$\tau$, since there will be transformations that leave $\Omega$ fixed
but change $\tau$.  In other words, the $\tau$ parameter is not faithfully
representing the duality group.  

In the rest of this paper we will find
the duality groups for general $SU(n)$ and describe  $\tau$
parameterizations that faithfully represents the duality group.  For the
$SU(3)$ case, there is a natural $\tau$ parameter to use.  For the higher
$SU(n)$ a parameterization can be found, but its relation to the actual
couplings, classical or otherwise, is less clear.

\newsec{Duality group for $SU(3)$}

This section is basically a review of arguments that have appeared 
elsewhere\refs{\MNI,\MNII}.
The group $SU(3)$ is a special case in that the duality group has a nonabelian
subgroup which is a subgroup of $SL(2,Z)$.  

Consider the case where all six hypermultiplet masses are zero.  Then the
Seiberg-Witten curve is given by \refs{\HO,\APS,\MNI}
\eqn\SUIIISW{
y^2=(x^3-ux-v)^2-(1-h^2(\tau))x^6
}
where $h(\tau)$ is a yet unspecified function of a coupling parameter $\tau$. 
The analysis is simplified if we consider the case where $u=0$.  The
Riemann surface is nonsingular, but it has a $Z_3$ symmetry, since
the curve is invariant under $x\to e^{2\pi i/3}x$.  Accordingly, $\Omega$
is invariant under a $Z_3$ Weyl group
transformation that is an element of the monodromy
group $Sp(4,Z)$.  

The $Sp(4,Z)$ transformation matrices are given
by $\left(\matrix{A &B\cr C &D}\right)$, where $A^TC-C^TA=B^TD-D^TB=0$
and $A^TD-B^TC=I$.  If $\Omega$ is given by
\eqn\Omeq{
\Omega=\tt \CC=\tt \left(\matrix{2&-1\cr-1&2}\right)
}
then it is invariant under the $Sp(4,Z)$ transformation 
$\Omega\to V\Omega V^T$,
where
\eqn\Veq{
V=\left(\matrix{0&1\cr-1&-1}\right),\qquad\qquad V^3=I
}
The matrix $\CC$ in \Omeq\ is the $SU(3)$ cartan matrix and we have
put a tilde over $\tau$ to distinguish it from the argument of $h(\tau)$.

We now examine what happens to $\Omega$ as we vary $\tau$ in $h(\tau)$.
Since the $Z_3$ symmetry is maintained as we continuously change $\tau$,
$\Omega$ remains proportional to the cartan matrix $\CC$.  Hence, up
to a $Z_3$ transformation, the generator for $\cT$ is given by
\eqn\Tgen{
\cT=\left(\matrix{I&\CC\cr0&I}\right)
}
It is possible that $\cT$ is  multiplied by some power of the 
$Z_3$ tranformation
$\cV=\left(\matrix{V&0\cr0&(V^{-1})^T}\right)$, which commutes with the
matrix in \Tgen.  
This $Z_3$ transformation is generated by $v\to e^{2\pi i}v$,
but such a term won't effect the argument presented here.

By the same argument, if we smoothly vary $h$ to $-h$, $\Omega$ remains
proportional to $\CC$, hence the $\cS$ transformation should  leave $\Omega$
proportional to $\CC$.  Up to a $Z_3$ transformation, the $\cS$ generator
is given by 
\eqn\Sgen{
\cS=\left(\matrix{0&-\sigma^{-1}\cr \sigma&0}\right)\qquad\qquad
\sigma=\left(\matrix{1&1\cr1&0}\right)
}
Under this transformation, $\tt$ in \Omeq\ is taken to $-1/3\tt$.

If we now turn $u$ back on, the duality transformations in \Tgen\ and
\Sgen\ cannot change since they are discrete.  Hence, these are the
generators of the duality group even when the $Z_3$ symmetry is broken.

Consider the transformation $\cS\cT\cS^{-1}$.  
This is given by 
\eqn\STSgen{
\cS\cT\cS^{-1}=\left(\matrix{I&0\cr-\sigma \CC\sigma& I}\right)=
\left(\matrix{I&0\cr 3\CC^{-1}&I}\right)
}
The $Z_3$ transformations commute with this generator as well.  It
is now a simple exercise to confirm the relation
\eqn\suIIIrel{
(\cS\cT\cS^{-1}\cT)^3=1.
}  
The $Z_3$ ambiguity is not important, since 
$\cS\cV^m\cT\cV^n\cV^{-m}\cS^{-1}\cT\cV^n=\cS\cT\cS^{-1}\cT\cV^{-n}\cV^n
=\cS\cT\cS^{-1}\cT$.

The generators $\cT$ and $\cS\cT\cS^{-1}$ generate a group that is
isomorphic to a subgroup of $SL(2,Z)$.  The action on $\tt$ is
\eqn\genact{
\cT:\ \ \tt\to \tt+1,\qquad\qquad \cS\cT\cS^{-1}:\ \ \tt\to {\tt\over -3\tt+1}
}
These are generators for the subgroup $\Gamma_1(3)$.  Hence, there
should exist a modular function $h(\tau)$ that bijectively maps the 
fundamental domain of $\Gamma_1(3)$ to the sphere and that also
has the correct weak coupling behavior.  Figure 1 shows the fundamental 
domain for $\Gamma_1(3)$, which is
four copies of the fundamental domain of $SL(2,Z)$. There is a 
$Z_3$ orbifold point
at $\tau=\omega=1/2+i/(2\sqrt{3})$, plus  cusps at $\tau=0$ and $\tau=\infty$.
The zeros of modular forms of weight $k$ satisfy the 
equation\refs{\Koblitz,\Lehner}
\eqn\zeroeq{
{v_\omega\over3}+v_0+v_\infty+\sum_{\{P\}}v_{\{P\}}= {k\over3}
}
where $v_{\{P\}}$ is the order of the zero at point $\{P\}$.  This
equation is 
similar to
the equation for modular forms of the full $SL(2,Z)$ group.
Clearly, to make a modular function of weight zero with a single zero and
pole, we should divide one weight three form by another weight three form.
The space of weight three forms is two dimensional and is generated by
$f_{\pm}(\tau)$, where\refs{\Koblitz,\MNII}
\eqn\fpmeq{
f_{\pm}(\tau)=\left({\eta^3(\tau)\over\eta(3\tau)}\right)^3\pm
\left(3{\eta^3(3\tau)\over\eta(\tau)}\right)^3.
}
 The desired function with the correct behavior is 
\eqn\hcorr{
h(\tau)=f_+(\tau)/f_-(\tau).
}
Using the almost modular properties of
$\eta(\tau)$, $\eta(-1/\tau)=(-i\tau)^{1/2}\eta(\tau)$, one readily finds
that $h(-1/(3\tau))=-h(\tau)$, hence $h$ has the correct transformation
under $\cS$.
$h(\tau)$ has a zero at $\omega$, where $f_+$ is
zero.  This is an Argyres-Douglas point\AD.  From \zeroeq\ we see that
$f_+$ is the cube of a weight 1 form.  There is also a pole at
$\tau=i/\sqrt{3}$, where $f_-(\tau)$ is zero.
\goodbreak\midinsert
\centerline{\epsfysize3in\epsfbox{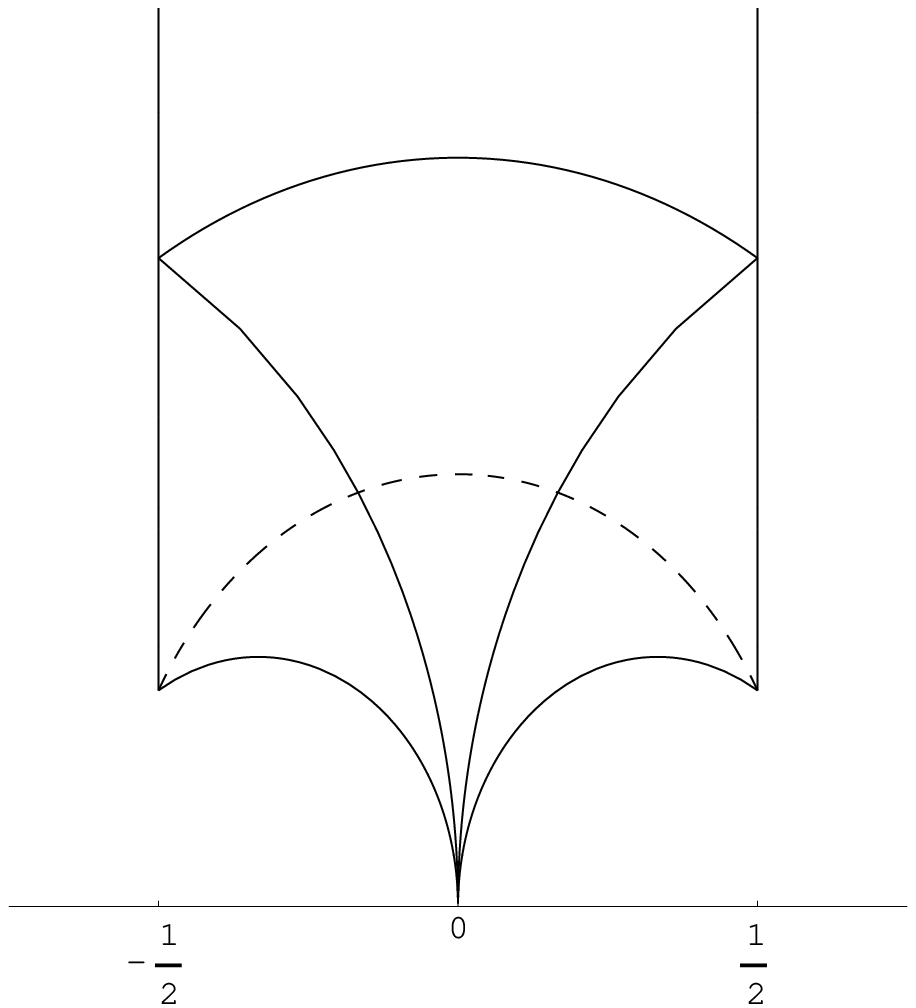}}
\leftskip 2pc
\rightskip 2pc\noindent{\ninepoint\sl \baselineskip=8pt {\bf Fig.~1}:
The fundamental region for $\Gamma_1(3)$, which is four copies of the 
$SL(2,Z)$ fundamental region.  The region below the dotted line is identified
with the region above the line under the $Z_2$ transformation 
$\tau\to -1/(3\tau)$.}
\endinsert

Using the functions in \fpmeq\ and \hcorr, one finds that for $u=0$,
the period matrix is $\tau C$.
In other words, the natural $\tau$ parameter that transforms faithfully
under the duality group is the $u=0$ effective coupling.  

\newsec{Duality groups for $SU(2n+1)$}

The duality group for  $SU(n)$ with $n>3$ does not have a nonabelian
subgroup that is a subgroup of $SL(2,Z)$.  In the case of of odd $n$,
the analysis for finding the duality group is 
similar to the $SU(3)$ case.

Suppose that $N_f=2n$ and all bare masses are zero.  Consider
the case where all casimirs are zero, except for the highest casimir
$u_n$.   The hyperelliptic curve then has the form
\eqn\curven{
y^2=(x^n-u_n)^2-(1-h^2(\tau))x^{2n}
}
This curve is invariant under a cyclic group of order $n$. Hence there
is an $Sp(2n-2,Z)$ transformation of order $n$ that leaves the 
period matrix $\Omega$ invariant.  A generator of this transformation
is $\left(\matrix{V&0\cr 0&(V^T)^{-1}}\right)$, where $V$ is an $n-1$
square matrix of the form
\eqn\Veq{
V=\left(\matrix{0&1&0&\dots&0\cr
                0&0&1&\dots&0\cr
                \vdots&&&\ddots\cr
		0&\dots&0&0&1\cr
		-1&-1&\dots&-1&-1}\right)
}
The appropriate period matrix $\Omega$ then satisifies 
\eqn\Omeq{
\Omega=V\Omega V^T.}

For example, in the $SU(5)$ case, the general matrix $\Omega$ satisfying 
\Omeq\ has the form
\eqn\OmsuV{
\Omega=	\left(\matrix{	2a&-b&b-a&b-a\cr
			-b&2a&-b&b-a\cr
			b-a&-b&2a&-b\cr
			b-a&b-a&-b&2a}\right)
}
Notice that the Cartan matrix $\CC$ has this form, but this does not
correspond to a hyperelliptic surface, basically because $\CC$ has
too much symmetry.  For a given $a$ there is a value $b$ such that
$\Omega$ is the period matrix for a hyperelliptic surface.  For what
follows, we will not need to know the actual form of $\Omega$, although
we will come back to it when we continue the discussion of the 
$SU(5)$ case.

The duality group is still generated by two elements $\cT$ and $\cS$.
Under $\cT$, $\Omega\to\Omega+\CC$, hence the generator is given by
\eqn\Tsuneq{
\cT=\left(\matrix{I&\CC\cr
		  0&I}\right).
}
Since $\CC=V\CC V^T$, this transformation
preserves the $Z_n$ symmetry.  The generator for the $\cS$ transformation
should have the form
\eqn\Ssuneq{
\cS=\left(\matrix{0&-(\sigma^T)^{-1}\cr
		 \sigma &0}\right).
}
Hence, if we smoothly change $\tau$ in \curven\ such that $h\to -h$, then
$\Omega\to -(\sigma^T)^{-1}\Omega^{-1}\sigma^{-1}$.  
The curve is still invariant under
the $Z_n$ symmetry, hence 
\eqn\sigeq{
V^T\sigma=\sigma (V)^m
}
 where $m$ is some
integer.  Since $\cS$ is an element of $Sp(2n-2,Z)$, $\sigma$ has integer
entries and $\det(\sigma)=\pm1$.  
If $m=1$, there exists a solution to \sigeq\ that satisfies the $Sp(2n-2,Z)$
requirements.  Up to a $Z_n$ transformation, this solution for
$\sigma$ is given by
\eqn\sigmaeq{
\sigma=\left(\matrix{1&\dots&1&1&1\cr
		     1&\dots&1&1&0\cr
		     1&\dots&1&0&0\cr
		     \vdots&&&&\vdots\cr
		     1&0&0&\dots&0}\right).
}

We now consider the combination of generators $\cS\cT\cS^{-1}\cT$, which
is given by
\eqn\combeq{
\cS\cT\cS^{-1}\cT=\left(\matrix{I&\CC\cr
  -\sigma \CC\sigma&I-\sigma \CC\sigma \CC}\right)=
\left(\matrix{I&\CC\cr-t\CC^{-1}&I-t}\right),
}
where $t=\sigma \CC\sigma \CC$.  It is then a simple exercise to
show that 
\eqn\combeqm{
(\cS\cT\cS^{-1}\cT)^m=
\left(\matrix{\CC f_m(t)\CC^{-1}&\CC g_m(t)\cr tg_m(t)\CC^{-1}&f_m(t)+tg_m(t)}
\right),
}
where
\eqn\fmgmeq{\eqalign{
f_m(t)&=\sum_{j=0}^{m-1}(-t)^j\left({m-1+j\atop m-1-j}\right)\cr
g_m(t)&=\sum_{j=0}^{m-1}(-t)^j\left({m+j\atop m-1-j}\right).
}}
As it so happens $g_m(x)$ is the characteristic polynomial for the
$SU(m)$ Cartan matrix.

Now let $m=n$ and suppose that $n$ is odd.  In this case the functions
$g_n(x)$ and $f_n(x)-1$ each factor into the product of two order $(n-1)/2$
 polynomials with integer coefficients.  For $g_n(x)$, we
have 
\eqn\gneq{\eqalign{
g_n(x)&=P_n(x)Q_n(x)\cr
P_n(x)&=n\sum_{j=0}^{(n-1)/2}(-x)^{(n-1)/2-j}{1\over n-2j}
\left({n-1-j\atop j}\right)\cr
Q_n(x)&=\sum_{j=0}^{(n-1)/2}(-x)^{(n-1)/2-j}
\left({n-1-j\atop j}\right),
}}
while for $f_n(x)-1$, the factorization is
\eqn\fneq{\eqalign{
f_n(x)-1&=P_n(x)R_n(x)\cr
R_n(x)&=-xg_{n-1\over2}(x).
}}
One can then show by induction that the characteristic polynomial for
$t$ when $n$ is odd is $(P_n(x))^2$.  Thus, from \gneq\ and \fneq\
one immediately concludes that $g_n(t)=f_n(t)-1=0$.  Therefore, the
generators $\cS$ and $\cT$ satisfy the identity
\eqn\STid{
\left(\cS\cT\cS^{-1}\cT\right)^n=1
\qquad\qquad{\rm if}\ n\ {\rm odd.}
}

If $n$ is even then \STid\ does not hold; it is obviously not true
for $SU(2)$.   Instead one finds that
\eqn\STide{
\left(\cS\cT\cS^{-1}\cT\right)^n=1+4\CD
\qquad\qquad{\rm if}\ n\ {\rm even,}
}
where $\CD$ is a matrix whose entries are either $\pm1$ or $0$ and
which satisfies $\CD^2=0$.  For example, for $SU(2)$, 
$\CD=\left(\matrix{-1&-1\cr1&1}\right)$.  Thus, for even $n$, the duality
group is still isomorphic to $\Gamma_0(2)$.

The identity \STid\ does not exist for any nontrivial elements of  
$SL(2,Z)$ if $n>3$.  Hence, the group
generated by $\cT$ and $\cS\cT\cS^{-1}$, the group that
leaves $h(\tau)$ invariant,
is not isomorphic to any $SL(2,Z)$ subgroup.  Thus, it is not
possible to express $h(\tau)$ as a modular function of an $SL(2,Z)$
subgroup and still have the duality group act faithfully on $\tau$.

Instead, let us consider a discrete subgroup,  $\Gamma$, of $SL(2,R)$ and look
for elements which have order $n$.  The subgroup should also have
an element of order 2, $\cS$,  in addition to the $\cT$ element that shifts
$\tau$ by $1$.  These elements generate the group.  The generator 
$\cS$ has the form $\left(\matrix{0&1/\sqrt{\al}\cr-\sqrt{\al}&0}\right)$,
so 
\eqn\SLIIRrel{
\cS\cT\cS^{-1}\cT=\left(\matrix{1&1\cr -\al&1-\al}\right).}
We then look for an $\al$ such that the $(\cS\cT\cS^{-1}\cT)^n=1$.
But we have already solved this; $\al$ has to be a root of the
polynomial $P_n(x)$ in \gneq.  If $\al$ is a root of $Q_n(x)$, then
it is still true that $(\cS\cT\cS^{-1}\cT)^n=1$, but there is also the
unwanted relation $(\cS\cT)^n=-1$.  

The factorization of $P_n(x)$ and $Q_n(x)$ is
\eqn\factor{
P_n(x)=\prod_{j=1}^{(n-1)/2}
\left(x-4\cos^2\left({\pi\over2}{2j-1\over n}\right)\right)\qquad\qquad
Q_n(x)=\prod_{j=1}^{(n-1)/2}
\left(x-4\cos^2\left({\pi\over2}{2j\over n}\right)\right)
}
The only root that is allowed is $\al=4\cos^2\left({\pi\over2n}\right)$.
A sketch of the proof is as follows\Lehner:  
Consider the transformation 
\eqn\trans{
W=\cS\cT^{-1}\cS\cT^{-1}\cS\cT^{-1}\cS=
\left(\matrix{1-\al&-1\cr2\al-\al^2&1-\al}\right)
}
This transformation has a fixed point in the upper half plane if $0<\al<2$.
But for these range of values and $\al\ne 1$ there is
no integer $q$ such that $W^q=1$, as can be easily verified by computing
the eigenvalues for $W$.  Therefore, every arc emanating out of the fixed
point is identitified with infinitely many arcs emanating out of the
same point.  Hence, for these values of $\al$, we see that $\Gamma$ is
not a proper discontinuous group.  Therefore, 
$\al=4\cos^2\left({\pi m\over2n}\right)$ is not allowed if $m>n/2$.
Likewise, we may consider all transformations of the form
\eqn\Vtrans{
V_m=(\cT\cS\cT\cS)^m\cT(\cS\cT\cS\cT)^m
}
If $m=1$, then there is a fixed point at 
$\tau=\sqrt{(\al-1)(\al-3)\over\al(\al-2)}$  Hence, this eliminates all values
in the range $2<\al<3$, since for these values the fixed point is in the upper 
half plane, yet there is no positive integer $q$ such that $(V_1)^q=1$.
A similar argument for the other values of $m$ shows that all values of 
$\al$ are eliminated
except for $\al=4\cos^2\left({\pi\over2n}\right)$ and 
$\al=4\cos^2\left({\pi\over n}\right)$.   However, the second value
has the relation $(\cS\cT)^n=-1$, so we discard it.

Let us consider the subgroup of $\Gamma$, $\Gamma'$, which is generated
by $\cT$ and $\cS\cT\cS^{-1}$.  The fundamental region lies within
the strip $-1/2\le\Re(\tau)\le1/2$.  There is a fixed point
at $\tau=\omega=\left(-1+i\tan\left({\pi\over2n}\right)\right)/2$.  
The transformation
$(\cS\cT\cS^{-1}\cT)^{(n+1)/2}$ maps the line segment at $\Re(\tau)=-1/2$
between the fixed point and the real line to an arc starting at the fixed
point and ending at the origin.  Likewise, $(\cT\cS\cT\cS^{-1})^{(n-1)/2}$
maps the line segment at $\Re(\tau)=1/2$ to the reflection of this arc.
Hence, schematically, the fundamental region for $\Gamma'$ looks like
the fundamental region shown in Figure 1.  This has the topology of
a sphere.  Figure 2 shows an overlay of the fundamental region of $\Gamma'$
for $n=3,5,7$ and $\infty$. 

\goodbreak\midinsert
\centerline{\epsfysize3in\epsfbox{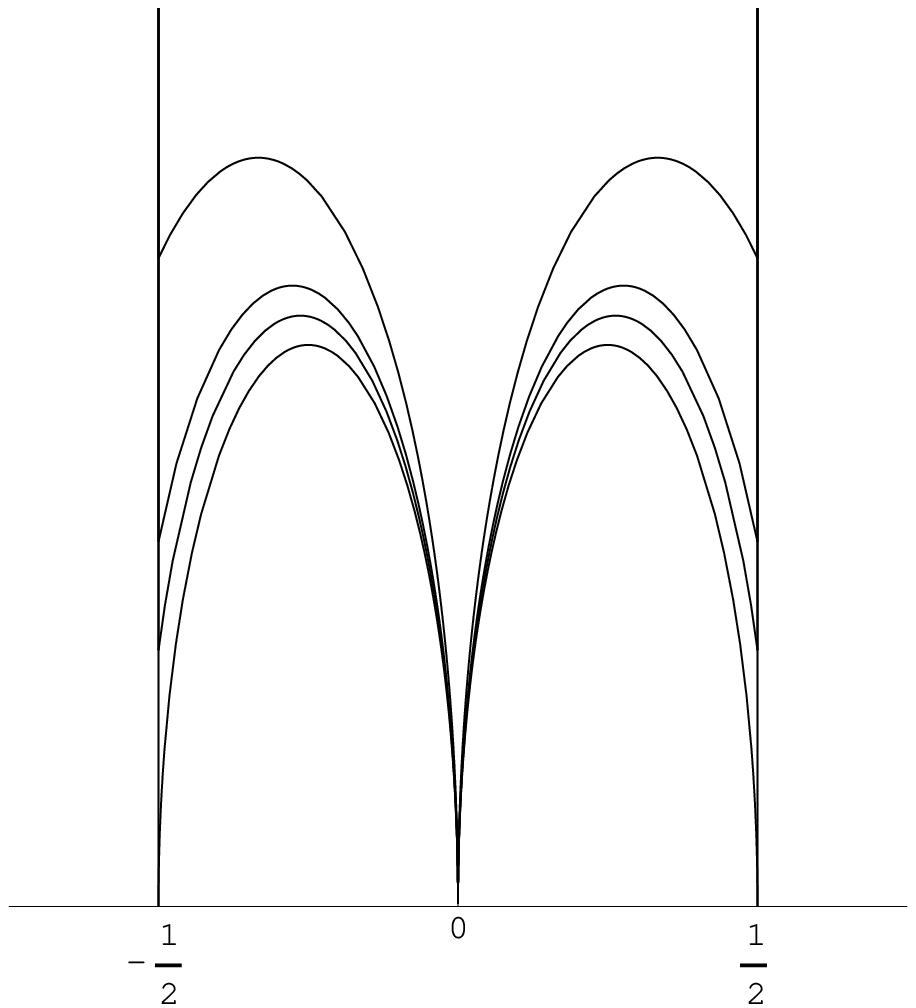}}
\leftskip 2pc
\rightskip 2pc\noindent{\ninepoint\sl \baselineskip=8pt {\bf Fig.~2}:
The fundamental regions for $\Gamma'$ when $n=3,5,7$ and $\infty$.}
\endinsert

The fundamental region for $\Gamma$ is found by identifying the region
of $\Gamma'$ inside the circle of radius $\sec\left({\pi\over2n}\right)$
with the region outside this circle.  

\newsec{Automorphic forms for higher $SU(n)$}

We now look for automorphic forms for the group $\Gamma'$.  Given
an element $\left(\matrix{a&b\cr c&d}\right)$, an automorphic
form with weight $k$ transforms as $F_k\left({a\tau+b\over c\tau+d}\right)=
(c\tau+d)^kF_k(\tau)$.  These automorphic forms satisfy an equation
similar to \zeroeq,
\eqn\zeroneq{
{v_\omega\over n}+v_0+v_\infty+\sum_{\{P\}}v_{\{P\}}= {(n-1)k\over2n}
}

For integer $k$ with $k\ge3$, consider the Poincar\'e series
\eqn\Poincare{
G_k(\tau)=\sum_{\{c,d\}\in \Gamma'}(c\tau+d)^{-k},
}
where the sum is over all distinct pairs $\{c,d\}$ that occur in $\Gamma'$.
This sum converges and has the transformation property
$G_k\left({a\tau+b\over c\tau +d}\right)=(c\tau+d)^kG_k(\tau)$ if
$\left(\matrix{a&b\cr c&d}\right)\in \Gamma'$, 
hence this is an automorphic form \Lehner.
For a given pair $\{c,d\}$, if $c=0$ then $d=1$.  Therefore, as 
$\tau\to i\infty$, $G_k(\tau)\to 1$.
We can also find another automorphic form by transforming the argument
of $G_k(\tau)$ by $\cS$, which is an element not in $\Gamma'$.  The resulting
form is 
\eqn\Heq{
H_k(\tau)=(-i\sqrt{\al}\tau)^{-k}G_k(-1/(\al\tau)).
}  
Since $d\ne 0$ for any pair $\{c,d\}$,  as $\tau\to i\infty$, 
$H_k(\tau)\to 0$.  The forms
$G_3(\tau)$ and $H_3(\tau)$ generate all possible automorphic forms of
weight 3.  In \zeroneq, we see that any form with weight 3 will have
a zero at one of the cusps,  a zero in the interior, or a zero of
order $3(n-1)/2$ at the orbifold point.  Hence there is a degree of 
freedom to move the position of this zero in the fundamental region,
therefore there are two independent automorphic forms of weight 3.
 
In particular, consider the two automorphic forms
\eqn\Fpmeq{
F_\pm(\tau)=G_3(\tau)\pm H_3(\tau).
}
Under $\cS$, we have
\eqn\Ftran{
F_\pm\left(-{1\over\al\tau}\right)=\pm i (\sqrt{\al}\tau)^3F_\pm(\tau)
}
If $\tau=\rho=i/\sqrt{\al}$, the fixed point of $\cS$, then we have that
\eqn\Fpmfp{
F_\pm(\rho)=\pm F_\pm(\rho).
}
Therefore $F_-(\tau)$ has a zero at $\rho$.  
Hence, $F_-(\tau)$ only has a zero of order $(n-3)/2$ at the orbifold
point $\omega$.  Since $F_+(\tau)$ has a zero of at least order $(n-3)/2$ at 
$\tau=\omega$, then
the automorphic function $F_+(\tau)/F_-(\tau)$ is nonsingular at
$\tau=\omega$.   However, $\omega$ is a fixed point under the transformation
$\cT^{-1}\cS$.  Hence, we have that
\eqn\Fpmom{
{F_+(\omega)\over F_-(\omega)}=- {F_+(\omega)\over F_-(\omega)}
}
Therefore, $F_+(\tau)$ must have a zero of order $3(n-1)/2$ at $\tau=\omega$,
and in fact $F_+(\tau)=(f_1(\tau))^{3(n-1)/2}$, where $f_1(\tau)$ is a form of
weight $2/(n-1)$ with a single zero at $\omega$ and normalized such
that $f_1(\tau)\to 1$ as $\tau\to i\infty$. 

We can then define two other
automorphic forms $f_\pm(\tau)$ with weight $2n/(n-1)$, where 
\eqn\fpmeq{
f_+(\tau)=(f_1(\tau))^n\qquad\qquad 
f_-(\tau)={F_-(\tau)\over (f_1(\tau))^{(n-3)/2}}.
}
The ratio $F_+(\tau)/F_-(\tau)=f_+(\tau)/f_-(\tau)$ then has a single
pole at $\rho$ and a zero of order $n$ at $\omega$.  But since $\omega$
is an order $n$ orbifold point, this maps to a single zero on the
Riemann sphere, ${\IC}\cup\{\infty\}$.  
In fact, 
\eqn\heq{
h(\tau)=f_+(\tau)/f_-(\tau)
}
is a bijective
map of the fundamental region of $\Gamma'$ to the Riemann sphere, and
so is the analog of the $j$ function for $SL(2,Z)$.  

Since $h(\tau)$ is $\pm1$ at the cusps, zero at $\omega$
and is a bijective map onto the Riemann sphere, it is appropriate to
identify $h(\tau)$ with the function $h(\tau)$ that appears in \curven.
As $\tau$ approaches the cusps, the theory becomes weakly coupled.
If $\tau=\omega$, then the theory is strongly coupled with mutually
nonlocal light states.

Unlike the case for $SU(3)$ and $SU(2)$, it does not seem possible to
express the function $h(\tau)$ in terms of Dedekind $\eta$-functions for
general $SU(n)$.  However,
it is
still possible to compute the coefficients of the Fourier expansion to
very high accuracy.  Consider the identity\Koblitz
\eqn\sumid{
\sum_{n\in Z}(z+n)^{-k}=(-1)^{k-1}2\zeta(k){k\over B_k}\sum_{j=1}^\infty
j^{k-1}e^{2\pi ijz},\qquad\qquad\Im z>0
}
where $B_n$ are the Bernoulli numbers.  Since $B_3=0$, this is not very
useful for $G_3(\tau)$ and $H_3(\tau)$.  However,  it is clear from
\zeroneq, \fpmeq\ and \heq\ that
\eqn\heqa{
h(\tau)={G_4(\tau)+H_4(\tau)\over G_4(\tau)-H_4(\tau)}.
}
so we can use \sumid\ to compute the fourier coefficients of $G_4$ and
$H_4$, which we can use in turn to compute the fourier coefficients
of $h(\tau)$.  

To find the coefficients, we need to know all pairs $\{c,d\}$ in
\Poincare.  Given one pair $\{c,d\}$, we can generate  new pairs either by
acting on the right with $\cT$ or $\cW=\cS\cT\cS^{-1}$.  In the first case,
the new pair is $\{c,d+c\}$, while in the second it is $\{c-\al d,d\}$.
For $G_4(\tau)$, we have that
\eqn\Gfeq{\eqalign{
G_4(\tau)&=\sum_{\{c,d\}}(c\tau+d)^{-4}=\sum_{\{c,d\}}c^{-4}(\tau+d/c)^{-4}\cr
&={8\pi^4\over3}\sum_{\{c,d\}_\cT}c^{-4}\sum_{j=1}^\infty
j^3 q^j,
}}
where $q=e^{2\pi i\tau}$ and the pair $\{c,d\}_\cT$ refers to all inequivalent 
pairs under the $\cT$ transformation ($\{c,d+c\}_\cT\equiv\{c,d\}_\cT$.)
For $H_4(\tau)$, we have that
\eqn\Hfeq{\eqalign{
H_4(\tau)&=\sum_{\{c,d\}}\left(-{c\over\sqrt{\al}}+d\sqrt{\al}\tau\right)^{-4}
={1\over\al^2}\sum_{\{c,d\}}d^{-4}\left(\tau-{c\over\al d}\right)^{-4}\cr
&={8\pi^4\over3\al^2}\sum_{\{c,d\}_\cW}d^{-4}\sum_{j=1}^\infty
j^3 q^j,
}}
the pair $\{c,d\}_\cW$ refers to all inequivalent 
pairs under the $\cW$ transformation, hence \break
$\{c-\al d,d\}_\cW\equiv\{c,d\}_\cW$.

As we argued earlier, $\al$ is a solution to the  equation
$P_n(x)=0$, where $P_n(x)$ is an order $(n-1)/2$ polynomial. 
Therefore, all pairs  $\{c,d\}$ are elements of the ring of integers
adjoined with $(n-3)/2$ powers of $\al$.  However, not all such
elements of the ring are entries for $c$ and $d$.  Otherwise, the number of
elements with $|c|<c_0$ and $|d|<d_0$, where $c_0$ and $d_0$ are finite,
is infinite and the group $\Gamma'$ would not be discrete\Lehner.  

As an example, let us return to the $SU(5)$ case.  All pairs
$\{c,d\}$ have the form $\{n_1+n_2\al,m_1+m_2\al\}$.  Since the 
determinant of any matrix $\left(\matrix{a&b\cr c&d}\right)$ is $1$
and $\al^2=5\al-5$, it immediately follows that $n_1=0\ {\rm mod}\ 5$
and $m_1=1\ {\rm mod}\ 5$. It is also easy to see that the only
combinations of even and odd  for the numbers $(n_1,n_2;m_1,m_2)$
are $(e,e;o,e)$, $(e,e;o,e)$, $(e,o;o,e)$, $(e,o;o,o)$, $(o,o;o,o)$ and
$(o,o;e,e)$.  It is also possible to find other rules.  For instance,
 $n_1=0$ only if $m_1=1$,
in which case $n_2|m_2$.  
It is also true that for any $d$ of the form $d=5p+1-q\al$ with $p\ne0$, then
$|3p+1|\le|q|\le|7p-1|$ and for $c$ of the form $c=5a-b\al$,
then $|3a|\le|b|\le |6a|+1$ if $a$ is odd and $|3a|+1\le|b|\le|6a|+1$
if $a$ is even and nonzero.

Even within these rules not all possible combinations of $c$ and $d$ are
allowed, since after a $\cT$ or a $\cW$ transformation, we might generate
a pair that violates the rules.  If this is the case, then we must
eliminate the original pair.  For example, the pair $\{5-3\al,-4+2\al\}$
is allowed, but $\{5-5\al,-4+2\al\}$ is not allowed, since under two  $\cT$
transformations we would generate $\{5-5\al,6-8\al\}$, which violates the
rule for allowed values of $d$.

In order to estimate the fourier coefficients for $G_4(\tau)$, it is
clear from \Gfeq\ that we want to find the smaller values of $|c|$ and
then given $c$, find the allowed values for $d$.  Likewise, to
estimate the coefficients in $H_4(\tau)$, we want to find the small
values of $|d|$ and then, given $d$, find the allowed values for $c$.
We find that the lowest allowed values for $c$ are
\eqn\callowed{
\pm(0,\al,-5+3\al,2\al,3\al,-5+5\al,4\al,7\al-10,9\al-15,5\al,8\al-10,7\al-5,
6\al,9\al-10...)
}
while the lowest allowed values for $d$ are
\eqn\dallowed{\eqalign{
&(1,-4+2\al,1+\al,1+2\al,-9+5\al,-4+4\al,1+3\al,-9+6\al,-14+8\al
...)\qquad d>0\cr
&(1-\al,1-2\al,6-4\al,1-3\al,1-4\al,11-7\al,6-6\al,1-5\al,11-8\al
...)\qquad\ \ \ d<0
}}
Figure 3 shows a plot of allowed values for $d$.  We can make a rough estimate
as to the asymptotic  number of allowed values of $d$ between $X$ and $X+1$.
Given that $d=-5n+1+m\al$, we find that the largest value of $n$ that gives
a $d$ value between $X$ and $X+2\al$ is approximately
$n_{max}\approx X/(3\al-5)$, while the smallest value is $n_{min}\approx 
X/(7\al-5)$.  For each even (odd) value of $n$ between 
$n_{max}$ and $n_{min}$, there
is one (are two) value(s) of $d$  between $X$ and $X+2\al$. Hence, to
leading order in $X$, we find that the number of allowed values between
$X$ and $X+1$ is
\eqn\dens{
{1\over2\al}{3\over2}\left(n_{max}-n_{min}\right)\sim {3X\over5 (11\al-16)}
}
This guarantees convergence for the sum in \Hfeq.

\goodbreak\midinsert
\centerline{\epsfysize3in\epsfbox{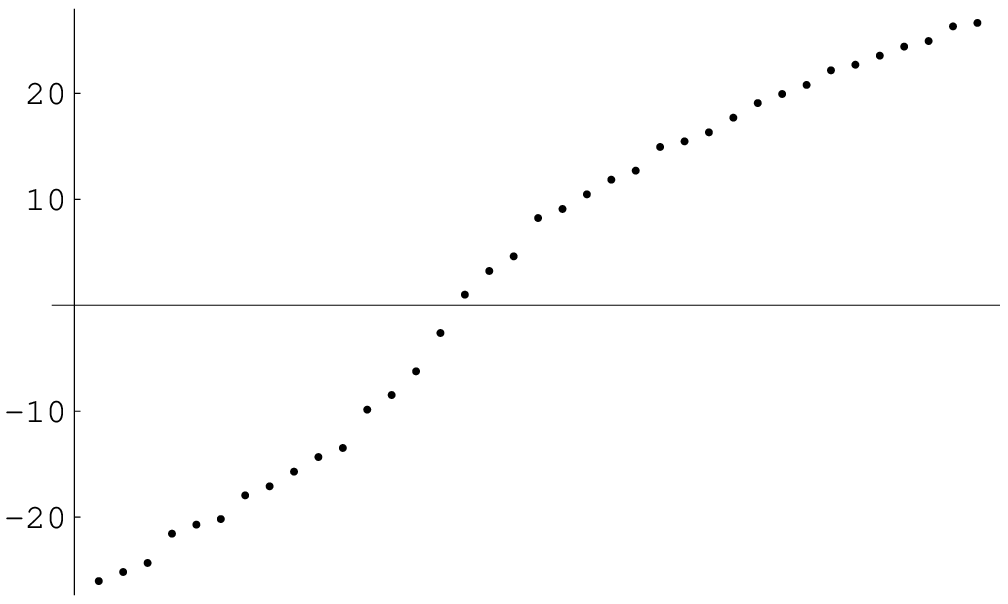}}
\leftskip 2pc
\rightskip 2pc\noindent{\ninepoint\sl \baselineskip=8pt {\bf Fig.~3}:
Allowed values for $d$. For large values of $d$, the density of allowed
values is roughly linear in $d$.}
\endinsert

We can now  estimate the coefficients of 
the $q$ expansion.  Using \sumid, \heqa, \Gfeq\ and \Hfeq\ 
the fourier expansion for $1-h^2(\tau)$ is approximately given by
\eqn\qexp{
1-h^2(\tau)=-\sum_{n=1}^\infty a_n q^n=
-76.2388q-3603.7q^2-140520. q^3-4.980 \times 10^6q^4+...
}
Compare this to the $n=3$ $q$ expansion where
\eqn\qexpIII{
1-h^2(\tau)=-108q-7128q^2-385884q^3-18950544q^4+...
} 
and the
$n=2$ expansion, where 
\eqn\qexpII{
1-h^2(\tau)=-64q-2560q^2-84736q^3-2551808q^4+...
}
The coefficients for the $n=5$ expansion lie between the coefficients
of these other two expansions. 
  
We still need a physical interpretation of the $\tau$ parameter.  The
bare coupling  $\tau_0$ is to leading order 
\eqn\taubare{
\tau_0={1\over2\pi i}\log((h^2(\tau)-1)/4)+{\rm O}\left(e^{2\pi i\tau}\right)
=\tau+{1\over2\pi i}\log(a_1/4)+{\rm O}\left(e^{2\pi i\tau}\right)
}
Inverting this gives 
\eqn\invert{
\tau=\tau_0-{1\over2\pi i}\log(a_1/4)+{\rm O}\left(e^{2\pi i\tau_0}\right)
}
The leading order
effective coupling in field theory for $u_i=0$, $i\ne n$ is
\eqn\coupeff{
\Omega=\tau_0 \CC + T
}
where $\CC$ is the Cartan matrix and the entries for $T$ are
\eqn\Tentries{\eqalign{
T_{mm}&={i\over2\pi}\log\left(16n^4\sin^4{\pi\over n}\right)
\qquad\qquad 1\le m\le n\cr
T_{m,m+1}=T_{m+1,m}&=-{i\over2\pi}\log\left(n^2
\tan^2{\pi\over n}\right)
\qquad\qquad 1\le m\le n-1\cr
T_{m,m+m'}=T_{m+m',m}=&{i\over2\pi}
\log\left(\cos^2{\pi\over n}-\cot^2{m'\pi\over n}\sin^2{\pi\over n}\right)^2
\qquad {1\le m\le n-2,\atop 2\le m'\le n-m}
}}

Letting $n=3$, one  finds that $T={i\over2\pi}\log(27)\CC$ and so
\eqn\OmegaIII{
\Omega=(\tau_0+\log(27))\CC=(\tau_0+\log(a_1/4))\CC.
}
Hence,  to leading order
$\tau$ is the bare coupling plus the quantum shift.

For $n=5$, we have that 
\eqn\TV{\eqalign{
T_{mm}&={i\over2\pi}2\log\left(25{5-\sqrt{5}\over2}\right)\qquad 
T_{m,m+1}=-{i\over2\pi}\log\left(25\left(5-2\sqrt{5}\right)\right)\cr
T_{m,m+m'}&=-{i\over2\pi}\log\left({1+\sqrt{5}}\right)
}}
so $T$ is not proportional to $\CC$.  If we compare the arguments of the
logs for $T_{mm}$ and $T_{m,m+1}$ with $a_1/4$, we see that the argument for 
$T_{mm}$ is greater than $a_1/4$, while the argument for $T_{m,m+1}$ is
less than $a_1/4$. Thus, the $\tau$ parameter lies somewhere between the 
quantum corrections for the diagonal and off-diagonal.  It is unclear what the
precise relation is.



\goodbreak
\vskip2.cm
\noindent{\it Note added:~~}After this paper was completed a paper appeared on
the archive that  overlaps with this one \AB.

\goodbreak
\vskip2.cm\centerline{\bf Acknowledgements}
\noindent
I thank D. Nemeschansky and N. Warner for helpful conversations and
E. Witten for an e-mail correspondence.
This work is supported in part
by funds provided by the DOE under grant number DE-FG03-84ER-40168.

\listrefs

\end